%% file: AAAI_main.tex
\documentclass[letterpaper]{article} 
\usepackage[]{aaai2027}  
\usepackage{times}  
\usepackage{helvet}  
\usepackage{courier}  
\usepackage[hyphens]{url}  
\usepackage{graphicx} 
\usepackage{natbib}  
\usepackage{caption} 
\usepackage{booktabs}
\usepackage{multirow}
\usepackage{subcaption}
\usepackage{graphicx}
\usepackage{amsmath,amssymb}
\input{macros.tex}
\usepackage{algorithm}
\usepackage{algorithmic}
\usepackage{dsfont}

\allowdisplaybreaks
\usepackage{newfloat}
\usepackage{listings}
\DeclareCaptionStyle{ruled}{labelfont=normalfont,labelsep=colon,strut=off} 
\floatstyle{ruled}
\newfloat{listing}{tb}{lst}{}
\floatname{listing}{Listing}
\title{Domain-Aware Machine Learning for Accelerating MILP-Based Motion Planning with Temporal Logic and Chance Constraints}
\author{
    Junyang Cai\textsuperscript{\rm 1},
    Weimin Huang\textsuperscript{\rm 1},
    Brendan Long\textsuperscript{\rm 2},
    Matthew Cleaveland\textsuperscript{\rm 2},
    Jyotirmoy V. Deshmukh\textsuperscript{\rm 1},
    Lars Lindemann\textsuperscript{\rm 3},
    Bistra Dilkina\textsuperscript{\rm 1}
}

\affiliations{
    \textsuperscript{\rm 1}
    Department of Computer Science,
    University of Southern California,
    Los Angeles, CA, USA\\

    \textsuperscript{\rm 2}
    MIT Lincoln Laboratory,
    Lexington, MA, USA\\

    \textsuperscript{\rm 3}
    Department of Information Technology and Electrical Engineering,
    ETH Zürich, Zürich, Switzerland\\

    caijunya@usc.edu,
    weiminhu@usc.edu,
    brendan.long@ll.mit.edu,
    matthew.cleaveland@ll.mit.edu,\\
    jdeshmuk@usc.edu,
    llindemann@ethz.ch,
    dilkina@usc.edu
}
\usepackage{bibentry}

\begin{document}

\maketitle

\begin{abstract}
Motion-planning problems with temporal-logic or chance constraints are often encoded as mixed-integer linear programs (MILPs). Although these encodings provide rigorous specifications, their combinatorial structure can make planning prohibitively slow. Machine learning for combinatorial optimization (ML4CO) has accelerated general-purpose MILP solving, but its standard graph representations discard semantic information available in control problems, such as variable roles, time indices, sample identities, and formula structure. We introduce a domain-aware ML4CO framework for MILP-based motion planning with temporal logic and chance constraints. The framework augments a conventional variable--constraint bipartite graph with features derived from the planning formulation and uses the resulting representation for two solver-guidance tasks: selecting branching backdoors and configuring solver parameters. We study three domains---Signal Temporal Logic (STL) planning, chance-constrained planning through Conformal Predictive Programming (CPP), and multi-agent Capability Temporal Logic (CaTL) planning---and compare against solver defaults, domain-agnostic learned methods, non-learned branching rules, MCTS transfer, and SMAC3 transfer. Across the tested distributions, domain-aware backdoor selection has the lowest reported mean solve time in all three domains, improving on default Gurobi by 14.4--20.4\%. Domain-aware configuration also has the lowest mean primal gap and primal integral in all three domains under a fixed SCIP time limit. These results show that exposing control-specific structure can improve learned MILP guidance beyond generic optimization features.
\end{abstract}

\section{Introduction}

Autonomous systems must satisfy complex, time-critical, and uncertain mission requirements. Examples include service robots that must visit target regions while avoiding obstacles and heterogeneous robot teams coordinating complementary capabilities in disaster-response missions. Such requirements are often expressed using temporal logics or chance constraints and compiled into mixed-integer linear programs (MILPs) \citep{bemporad1999control,raman2014model,belta2019formal,zhao2024conformal}. MILP formulations are attractive because mature solvers can find feasible plans and, given sufficient time, certify optimality. They have therefore been widely used in aircraft trajectory planning, vehicle routing, and robot motion planning \citep{richards2002aircraft,schouwenaars2001mixed,kamale2024optimal,ioan2021mixed}.

However, MILP-based planning is computationally challenging because the resulting problems are NP-hard \citep{karp2009reducibility}. Solve times can grow rapidly with the planning horizon, number of agents or uncertainty samples, and specification complexity. Even specialized encodings that reduce the number of binary variables \citep{kurtz2022mixed} may fail to meet the real-time requirements of autonomous systems.

\begin{figure*}[t]
    \centering
    \includegraphics[width=0.88\textwidth]{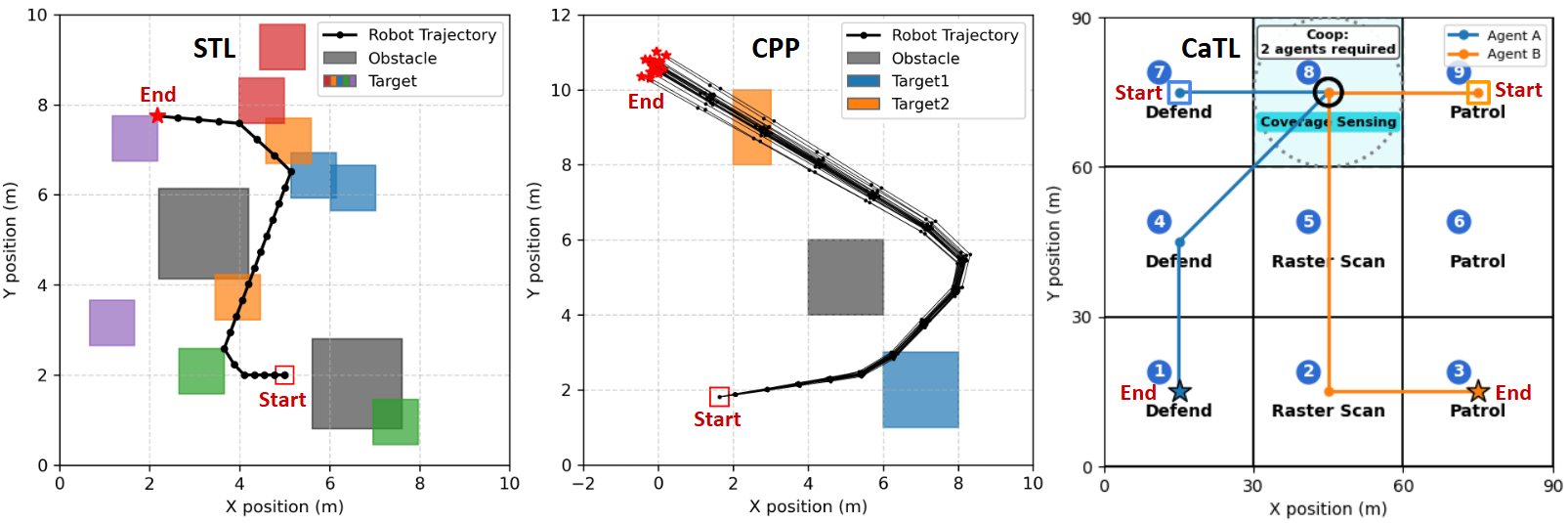}
    \caption{Representative plans from the three case studies: STL multi-target navigation with obstacle avoidance (left), CPP chance-constrained timed reach-avoid planning under sampled disturbances (middle), and CaTL coordination of heterogeneous agents with capability-dependent tasks (right).}
    \label{fig:robot-trajectory}
\end{figure*}

Machine learning for combinatorial optimization (ML4CO) offers a promising way to improve computational efficiency without replacing the symbolic solver. Graph Neural Networks (GNNs) can represent an MILP as a bipartite graph and learn to guide solver decisions such as branching, partial assignment, and solver configuration \citep{gasse2019exact,cai2024learning,scavuzzo2024machine}. Previous methods have reduced solve time and search-tree size and, in some cases, outperformed expert-designed heuristics \citep{huang2023searching,huang2024contrastive,cai2024multitaskrepresentationlearningmixed}. Most existing approaches, however, use only generic MILP features, such as coefficients, bounds, and linear-programming statistics, while ignoring the semantics of the model that generated the MILP.

This omission is particularly important in motion planning. Variables at adjacent time steps have related roles; binary variables may encode predicates or temporal operators; chance-constrained formulations repeat structure across uncertainty samples; and multi-agent formulations couple occupancy, capability, and temporal requirements. These semantic relationships may help a learned policy distinguish important solver decisions that appear similar from the coefficient matrix alone.

\begin{figure*}[t]
    \centering
    \includegraphics[width=0.82\textwidth]{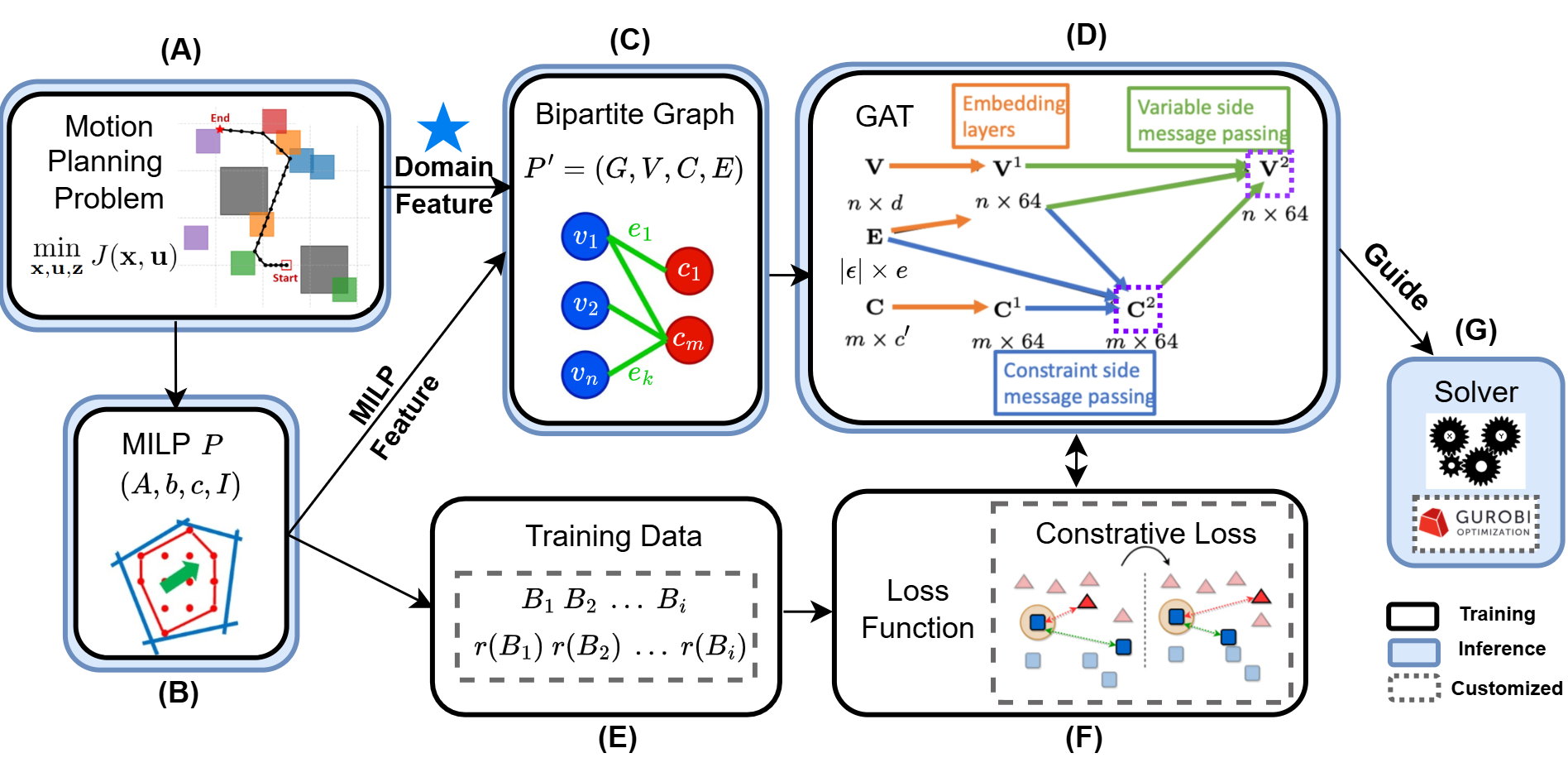}
    \caption{Overview of the proposed domain-aware ML4CO pipeline. A planning specification is compiled into an MILP and metadata sidecar. Generic MILP features are augmented with semantic roles and domain indices, embedded by a bipartite GAT, and passed to task-specific heads for backdoor ranking or solver configuration. The selected decision guides the unmodified solver.}
    \label{fig:pipeline}
\end{figure*}

We therefore ask: \emph{Can semantic information from formal planning models improve learned guidance of an off-the-shelf MILP solver?} We address this question through a domain-aware neuro-symbolic framework for two solver-guidance tasks: learned backdoor selection and instance-conditioned solver configuration. As illustrated in Fig.~\ref{fig:pipeline}, we augment a generic bipartite MILP representation with metadata from the planning model, including semantic roles, normalized time and sample indices, constraint types, and formula information. A GNN uses this representation to guide an otherwise unmodified MILP solver.

We study three representative planning domains: motion planning under Signal Temporal Logic (STL) specifications \citep{raman2014model}, chance-constrained planning reformulated through Conformal Predictive Programming (CPP) \citep{zhao2024conformal}, and multi-agent coordination under Capability Temporal Logic (CaTL). Representative trajectories are shown in Fig.~\ref{fig:robot-trajectory}.

Our contributions are:
\begin{itemize}
\item We develop a neuro-symbolic ML4CO framework that guides a conventional MILP solver without modifying the underlying planning formulation.
\item We introduce a domain-aware bipartite representation that combines generic MILP features with semantic roles, temporal and sample indices, constraint types, and formula information.
\item We evaluate the framework on backdoor selection and solver configuration across STL, CPP, and CaTL problems, comparing it with domain-agnostic variants, solver defaults, and non-learned and optimization-based baselines.
\end{itemize}

The learned components affect only search efficiency; feasibility and optimality guarantees remain properties of the MILP formulation and the underlying solver.

\section{Background and Problem Formulation}
We are interested in motion planning problems for discrete-time dynamical systems of the form
\begin{equation}
\label{eq:dynsys}
    \vx_{t+1} = f_t(\vx_t,\vu_t)
\end{equation}
where $t \in \{0,\ldots,T\}$ denotes time over a finite time horizon $T$,  $\vx_t\in\mathbb{R}^{n_\vx}$ and $\vu_t\in\mathbb{R}^{n_\vu}$ are the state and control input of the system at time $t$, respectively, and $f$ describes the system dynamics. We also introduce the notation $\vx:=(\vx_{0}, \hdots, \vx_{T})$ and $\vu:=(\vu_{0}, \hdots, \vu_{T-1})$. Finally, we assume that the initial system state $\vx_0$ lies
in some compact set $X_0$, i.e., $\vx_0 \in X_0$.

\paragraph{Representing and Solving MILPs} MILPs are optimization problems with discrete decision variables. Formally, MILPs are the class of problems with the form
\begin{equation}
\label{eq:milp}
\min\{c^Tx \mid Ax \leq b, \, x \in \mathbb{R}^n, \, x_j \in \{0,1\}, \, \forall j \in I\}, 
\end{equation}
where $c \in \mathbb{R}^n$, $b \in  \mathbb{R}^m$, and $A \in  \mathbb{R}^{m\times n}$. $I \subseteq \{1,...,n\}$ is the set of variables that are restricted to be integers. 

The Branch-and-Bound (BnB) algorithm is an exact tree search algorithm for solving MILPs, which is the core component in common MILP solvers. BnB breaks the original MILP down into smaller subproblems by splitting the domain of an integer variable and maintains upper and lower bounds to eliminate subproblems.

\paragraph{MILP-based Motion Planning} Mixed integer-based motion planning problems for the system $\vx_{t+1} = f(\vx_t,\vu_t)$ can generally be written as
\begin{subequations}\label{eq:milp_oc}
\begin{align}
    \min_{\vx,\vu,\vz} \quad & J(\vx,\vu)\\
    \textrm{s.t.} \quad & \vx_{t+1} = f_t(\vx_t,\vu_t) \; \text{ for all } t  \in \{0,\ldots,T-1\} \\
    & \vx_0\in X_0\\
    & c(\vx,\vz) = 1 \label{eq:binary_constraint}
\end{align}
\end{subequations}
where $\vz:=(\vz_{0}, \hdots, \vz_{T})$ is a set of auxiliary integer decision variables with $\vz_t\in\mathbb{Z}^{n_\vz}$ denoting the integer decision variables at time $t$.  Importantly, $c$ is a Boolean-valued function that depends on these   integer decision variables as well as the continuous states to encode the motion task in hand via $c(\vx,\vz) = 1$. In this paper, these tasks will be logic and chance-constrained tasks for which we derive the specific forms of the function $c$ in the subsequent sections. We also introduced a cost function $J$ and mention that we could easily incorporate state and input constraints for $\vx_t$ and $\vu_t$. Due to the existence of integer decision variables $\vz$ and if we restrict the functions $J$, $f$, and $c$ to be linear in their arguments, we can map the motion planning problem directly to the MILP with the decision variable $x:=(\vx,\vu,\vz)$ from which the definitions of $A$, $b$, $c$, and $I$ follow. 

\paragraph{Discrete-Time Signal Temporal Logic} 
Signal Temporal Logic (STL) \citep{maler2004monitoring} is a specification
language for expressing properties of  real-time signals. A signal $\vx$ is a
function from a time domain to a value domain. In this paper, the value domain is $\mathbb{R}^{n_\vx}$ (or subsets of $\mathbb{R}^{n_\vx}$).  Discrete-time STL (DT-STL) is a variant of STL where
the signal is defined over the discrete time domain $\{0,1,\ldots,T\}$ with $T$ being
the time horizon. Therefore, the signal $\vx$ may represent the trajectory of the system $\vx_{t+1} = f(\vx_t,\vu_t)$. The basic syntax of a DT-STL
formula is as follows:
\begin{equation}
\label{eq:stlfrag}
\varphi\  =\ \top \mid h(\vx_t) \ge 0 \mid \neg \varphi_1 \mid 
             \varphi_1 \wedge \varphi_2 \mid \varphi_1 \until_\intvl \varphi_2.
\end{equation}
Here, $\top$ is the Boolean true symbol and $h(\vx_t)\in \mathbb{R}$ is a function mapping $\vx_t$ to a real value. The symbols $\neg$ and $\wedge$ denote the standard Boolean
operators for negation and conjunction. The symbol  $\until_\intvl$  denotes the temporal operator for until that is defined over the time interval $\intvl\subseteq \mathbb{R}_{\ge 0}$. The formula $\varphi_1 \until_\intvl \varphi_2$ expresses that $\varphi_1$ is true until $\varphi_2$ becomes true within the interval $\intvl$. Additionally, we can derive Boolean operators such as disjunction (denoted by the symbol $\vee$) or temporal operators such as eventually and always (denoted by the symbols $\ev_\intvl$ and $\alw_\intvl$). The Boolean semantics of DT-STL extends the semantics of
propositional logic to timed traces. Specifically, by $(\vx,t)\models \varphi$ we denote that the signal $\vx$ satisfies the formula $\varphi$ at time $t$. Quantitative semantics instead define a score function $\rho^\varphi(\vx,t)\in\mathbb{R}$ that maps a formula $\varphi$, a signal $\vx$, and a time $t$ to a real value. We have that $\rho^\varphi(\vx,t)>0$ implies $(\vx,t)\models \varphi$. We omit definitions of Boolean and quantitative semantics and refer the  reader to \citep[Chapter 3]{lindemann2025formal}.

\begin{table*}[t]
\centering
\caption{Metadata attached to each MILP variable for STL, CPP, and CaTL. Each row shows the information retained from the original domain model before conversion to the generic MILP representation.}
\label{tab:features}
\begin{tabular}{@{}p{0.08\textwidth}p{0.40\textwidth}p{0.17\textwidth}p{0.25\textwidth}@{}}
\toprule
Domain & Variable role & Position in the model & Additional domain information\\
\midrule
STL
& boolean operator, predicate, system state, control input, output, robustness, or auxiliary variable
& timestep and depth in the STL formula tree
& target / obstacle distance\\
\midrule
CPP
& shared control, sample-specific state, sample-specific output, robustness, or auxiliary variable
& timestep and sample index
& sampled uncertainty\\
\midrule
CaTL
& control or occupancy variable, or indicator variable associated with a temporal, capability, proposition, conjunction, or disjunction constraint
& timestep and capability index
& region index\\
\bottomrule
\end{tabular}
\end{table*}

We say that a DT-STL formula is well-formed 
if the formula horizon is less than $T$, where the formula horizon is defined as: $\horizon(\top) =  0$, $\horizon(h(\vx_t)\ge 0) =  0 $, $\horizon(\neg \varphi_1)  =  0$, $\horizon(\varphi_1 \wedge \varphi_2)\  = 
    \max(\horizon(\varphi_1), \horizon(\varphi_2)) $, and $\horizon(\varphi_1 \until_I \varphi_2)\ 
 =  \max(I) +\max(\horizon(\varphi_1),   \horizon(\varphi_2)) $. A formula horizon less than $T$ guarantees that a well-formed
DT-STL formula can be unambiguously evaluated over the signal $\vx$ which is defined
over  $\{0,1,\ldots,T\}$.

\subsection{Encoding DT-STL Planning Problems as MILP-Based Planning Problems}
We are now in a position to formulate the 
DT-STL motion planning problem which differs from the MILP-based planning problem in \eqref{eq:milp_oc} only in the last constraint $c(\vx,\vz) = 1$, which is now replaced by the satisfaction constraint $(\vx,0) \models \varphi$. In our implementation, we use the stlpy package \citep{kurtz2022mixed} to encode the constraint $(\vx,0) \models \varphi$ into mixed-integer constraints, which assumes $\varphi$ is in negative normal form \citep[Section 5]{belta2019formal}. This is without loss of generality as every DT-STL formula $\varphi$ can be translated into a semantically equivalent DT-STL formula $\varphi_\text{NNF}$ that is in negative normal form \citep[Section 4]{sadraddini2015robust}. For this MILP encoding, predicates are encoded using binary variables via the Big-M method. Boolean and temporal operators are handled through recursive binary inequalities. Satisfaction of the constraint is enforced by setting the root binary variable $\vz_0^{\varphi_\text{NNF}}=1$, corresponding to integer constraint within the MILP-based planning problem \eqref{eq:milp_oc}. A detailed formulation of the DT-STL problem and the corresponding MILP encoding is provided in Appendix.

\subsection{Encoding Chance Constrained Planning Problems as MILP-Based Planning Problems}
Chance constrained programming is typically used when the dynamical system $\vx_{t+1} = f(\vx_t,\vu_t)$, the environment that the system operates in, or the task itself is uncertain.  To handle uncertainty, the  constraint $c(\vx,\vz) = 1$ in the MILP-based planning problem \eqref{eq:milp_oc} is replaced by the chance constraint $\text{Prob}(g(\vx,\vw)\ge 0)\ge 1-\delta$, where $\vw$ represents random variables describing uncertainty. Since evaluating the chance constraint analytically is often intractable, we apply Conformal Predictive Programming (CPP) \citep{zhao2024conformal} to approximate the chance constraint using $K$ samples  over which the constraint $\sum_{i = 1}^K \mathds{1} {g(\vx^{(i)},\vw^{(i)}) \ge 0} \ge \lceil(K + 1)(1-\delta)\rceil$ is enforced. The constraint denotes the empirical quantile over the distribution of sampled constraints and we can further reformulate this as a MILP by assigning a binary variable $\vz_i$ to each sample $i$ using the Big-M method.  The precise formulation and MILP encoding of chance constraints are shown in Appendix.

\subsection{Encoding CaTL Planning Problems as MILP-Based Planning Problems}

Capability temporal logic (CaTL) is another temporal logic -- similar to STL -- that is specifically designed for heterogeneous multi-agent planning in discrete environments \citep{leahy2021scalable}. The CaTL planning problem therefore replaces the constraint $c(\vx,\vz) = 1$ in the MILP-based planning problem \eqref{eq:milp_oc}  by the constraint $(\vx,0) \models \varphi$, where  $\varphi$ is now a CaTL formula.  CaTL planning problems are high-level planning problems where  the environment is partitioned into regions with interconnections and where each agent is represented solely by its capabilities (e.g. agents have different sensors). The CaTL atomic propositions, called {tasks}, count the number of agents with a specific capability that are in a specific region of the environment for a specified number of discrete time steps. For instance, consider the environment shown in the right panel of Figure \ref{fig:robot-trajectory} where a plan for the following requirement expressed as a CaTL formula is shown: (1) sense region 8 with capabilities A and B simultaneously, (2) defend region 1 with capability B, (3) patrol region 3 with capability A, and (4) maintain one agent in region 1 until region 3 is reached. For a more detailed description of the CaTL planning problem, we refer the reader to Appendix.

\section{Domain-Aware ML-Guided Solving}
Figure~\ref{fig:pipeline} summarizes our framework. A planning model produces both an MILP and a metadata sidecar that records the semantic origin of its variables and constraints. We combine this metadata with generic MILP features, process the resulting graph with a GNN, and use its output to guide BnB.

\subsection{Domain-Aware Bipartite Representation}
Following \citet{gasse2019exact}, we represent an MILP with variables $x\in\mathbb R^n$ and constraints $Ax\leq b$ as a bipartite graph $\mathcal G=(\mathcal V,\mathcal C,\mathcal E)$. Variable node $i\in\mathcal V$ represents $x_i$, constraint node $j\in\mathcal C$ represents row $j$, and $(i,j)\in\mathcal E$ iff $A_{ji}\neq0$. Generic features comprise 15 variable features (including type, objective coefficient, bounds, and root-LP statistics), four constraint features (including right-hand side and sense), and the edge coefficient.

In addition to the generic MILP features derived from $(A,b,c)$, we attach a metadata vector to each decision variable. The metadata describe the modeling role of the variable and its position in the original planning formulation. For every variable, the final feature vector contains three groups of information: (i) a one-hot encoding of its variable role, (ii) normalized structural indices such as time, formula depth, sample index, or capability index, and (iii) domain-specific scalar quantities. Categorical attributes are one-hot encoded, while integer-valued indices are divided by the largest corresponding index in the same instance. Features that do not apply to a particular domain or variable are set to zero. Table~\ref{tab:features} summarizes the domain-specific schema.

\begin{table*}[!t]
\centering
\caption{Benchmark statistics for the two solver-guidance tasks. Parameters are (obstacles, target groups, targets/group, horizon) for STL; (samples, horizon) for CPP; and (capability labels, regions, horizon) for CaTL. Variable counts are reported as (binary, integer, continuous).}
\label{tab:benchmarks}
\begin{tabular}{@{}lllrrr@{}}
\toprule
Task & Benchmark & Parameters & Vars. (B, I, C) & Constraints & Default Gurobi (s)\\
\midrule
\multirow{3}{*}{Backdoor selection}
 & STL  & $(2,5,2,30)$ & $(2{,}801,0,435)$ & 4,605 & 162.5\\
 & CPP  & $(20,15)$ & $(2{,}320,740,60)$ & 15,201 & 228.3\\
 & CaTL & $(5,5,42)$ & $(5{,}353,5{,}819,1)$ & 22,960 & 59.5\\
\midrule
\multirow{3}{*}{Solver configuration}
 & STL  & $(3,5,3,35)$ & $(1{,}728,0,3{,}756)$ & 7,073 & 191.6\\
 & CPP       & $(20,15)$ & $(2{,}320,740,60)$ & 15,201 & 228.3\\
 & CaTL & $(5,5,50)$ & $(12{,}207,7{,}275,1)$ & 47,586 & 746.9\\
\bottomrule
\end{tabular}
\end{table*}

The network first projects variable, constraint, and edge features into a common $L$-dimensional space. Let $h_i^V$, $h_j^C$, and $h_{ij}^E$ denote the resulting embeddings. In the first message-passing round, every constraint attends to its incident variables,
\begin{equation}
 h_j^{C'}=\operatorname{GAT}_C\!\left(h_j^C,
 \{(h_i^V,h_{ij}^E):(i,j)\in\mathcal E\}\right). \nonumber
\end{equation} 
A second round reverses the direction and updates each variable from its incident constraints,
\begin{equation}
 h_i^{V'}=\operatorname{GAT}_V\!\left(h_i^V,
 \{(h_j^{C'},h_{ij}^E):(i,j)\in\mathcal E\}\right). \nonumber
\end{equation}
Multi-head attention allows the model to weight different constraint relationships without imposing an ordering on variables or rows. Separate task heads produce the outputs described next. Domain-aware and generic variants use the same architecture, optimization, data split, and random seed; only the metadata block differs. This controlled design is important: improvements over the generic variant cannot be attributed to a deeper network, a different loss, or additional training instances.

\subsection{Backdoor Selection}
A backdoor is a small set of integer variables that, when prioritized for branching, can produce a smaller BnB tree \cite{dilkina2009backdoors}. Following \citet{ferber2022learning}, we learn to rank candidate backdoor sets. For an instance $P$ and candidate set $B_i$, the model outputs $s_i=\pi(P,B_i;\theta)$, where a larger score denotes a better candidate. For a pair $(B_1,B_2)$ with measured runtimes $r(B_1)$ and $r(B_2)$, let $y=1$ if $r(B_1)<r(B_2)$ and $y=-1$ otherwise. We minimize
\begin{equation}
\label{eq:ranking}
 \ell(s_1,s_2,y)=\max\{0,-y(s_1-s_2)+m\},
\end{equation}
where $m>0$ is a margin. This convention makes the training objective consistent with selecting the highest-scoring candidate at inference.

We use MCTS \cite{khalil2022finding} to generate candidates for each training instance, retain the 15 fastest and 15 slowest sets, and form ranking pairs between them. At test time, we sample 50 size-eight candidates from fractional variables in the root LP, score them, and assign the highest-ranked set elevated branching priority. \textbf{Backdoor-Rank} uses generic graph features; \textbf{DA-Backdoor} adds our domain-aware features.

\subsection{Instance-Conditioned Solver Configuration}
MILP solvers expose parameters controlling branching, cuts, heuristics, and LP processing. We learn an instance-conditioned configuration over 15 SCIP parameters. SMAC3 \cite{lindauer2022smac3} generates candidate configurations for each training instance. The best feasible configurations within a fixed labeling budget form a positive set $S_+^P$, and the worst form a negative set $S_-^P$. A graph-level task head emits an embedding $\pi(P;\theta)$, trained with the multi-positive contrastive objective
\begin{equation}
\label{eq:contrastive}
-\sum_P\frac{1}{|S_+^P|}\sum_{a\in S_+^P}
\log\frac{\exp(a^\top\pi(P;\theta)/\tau)}
{\sum_{a'\in S_-^P\cup\{a\}}\exp(a'^\top\pi(P;\theta)/\tau)}, \nonumber
\end{equation}
where $\tau$ is a temperature. At inference, the task head is decoded into valid parameter values and applied before solving. \textbf{Config-CL} uses generic features, whereas \textbf{DA-Config} uses the augmented representation.

\begin{table*}[!t]
\centering
\caption{{\textbf{Backdoor selection:} Results are measured on 100 test instances per domain with a 900-second cutoff. ``Wins'' are exclusive per-instance minima as reported in the experiment records. Parentheses show mean-runtime speedup relative to Default Gurobi; negative values denote slowdowns. Lower runtime is better, and best values for each domain are in bold.}}
\label{tab:backdoor}
\begin{tabular}{llrrrrrr}
\toprule
Benchmark & Method & Wins & Mean (s) & Std. Dev. & 25th pct. & Median & 75th pct.\\
\midrule
\multirow{6}{*}{STL}
& Default Gurobi & 10 & 162.5 & 231.8 & 21.0 & 77.5 & 177.7\\
& Random & \textbf{34} & 145.1 (10.7\%) & 217.3 & \textbf{18.5} & 65.4 & 148.0\\
& LP-Frac & 12 & 154.2 (5.1\%) & 225.0 & 19.7 & 71.2 & 164.1\\
& MCTS-Transfer & 13 & 153.7 (5.4\%) & 214.4 & 22.4 & 73.2 & 151.8\\
& Backdoor-Rank & 16 & 145.0 (10.8\%) & 205.3 & 21.5 & 58.7 & 174.8\\
& \textbf{DA-Backdoor (ours)} & 12 & \textbf{129.9 (20.1\%)} & \textbf{203.8} & 20.3 & \textbf{55.4} & \textbf{127.5}\\
\midrule
\multirow{6}{*}{CPP}
& Default Gurobi & 20 & 228.3 & 156.3 & 119.6 & 178.1 & 288.5\\
& Random & 15 & 357.0 ($-56.4\%$) & 269.1 & 141.8 & 248.0 & 534.5\\
& LP-Frac & 20 & 331.9 ($-45.4\%$) & 249.6 & 154.5 & 260.2 & 408.0\\
& MCTS-Transfer & 6 & 368.4 ($-61.4\%$) & 185.8 & 230.5 & 326.8 & 457.4\\
& Backdoor-Rank & 13 & 202.4 (11.3\%) & 124.9 & \textbf{114.3} & \textbf{162.8} & 255.9\\
& \textbf{DA-Backdoor (ours)} & \textbf{22} & \textbf{195.5 (14.4\%)} & \textbf{114.0} & 114.7 & \textbf{162.8} & \textbf{255.1}\\
\midrule
\multirow{6}{*}{CaTL}
& Default Gurobi & 7 & 59.5 & 30.7 & 38.8 & 46.7 & 74.0\\
& Random & 6 & 68.6 ($-15.3\%$) & 35.8 & 42.3 & 57.9 & 82.5\\
& LP-Frac & 11 & 57.7 (3.1\%) & 28.0 & 37.0 & 47.5 & 71.4\\
& MCTS-Transfer & 27 & 54.8 (7.9\%) & 23.5 & 37.4 & 50.4 & 65.6\\
& Backdoor-Rank & 21 & 50.7 (14.7\%) & 23.2 & 37.4 & 44.3 & 56.5\\
& \textbf{DA-Backdoor (ours)} & \textbf{28} & \textbf{47.4 (20.4\%)} & \textbf{17.4} & \textbf{36.4} & \textbf{41.8} & \textbf{52.2}\\
\bottomrule
\end{tabular}
\end{table*}

\section{Problem Domain Description}

\paragraph{Multi-Target Signal Temporal Logic Planning}
We evaluate a specific STL task, namely the multi-target problem (see Fig.~\ref{fig:robot-trajectory}). In this setting, a robot must avoid obstacles (grey) and visit at least one target of each type (color). Such specifications frequently arise in real-world applications, e.g., mobile service robots or delivery drones operating in cluttered environments.
We consider a 2D mobile robot, where $f$ is modeled by double-integrator dynamics, that must navigate through a field of obstacles (denoted by $\mathcal{O}_i$) and reach at least one target of each color (denoted by $T_i^j$). Formally, the STL specification is given by
$
\varphi = \bigwedge_{i=1}^{N_c} \left( \bigvee_{j=1}^{N_t} F_{[0,T]} T_i^j \right) \land G_{[0,T]} \left( \bigwedge_{k=1}^{N_o} \neg \mathcal{O}_k \right).
$
The cost function $J=\sum_{t=0}^{T-1}\sum_{i=1}^{n_u} \vu_{t,i}$  penalizes the accumulated control effort over time.

\paragraph{Chance-Constrained Timed Reach-Avoid Planning} 

We consider a chance constrained planning problem where
a robot, where $f$ is again modeled with double integrator dynamcis, that is subject to disturbances $\vw$ has to reach two successive goal regions ($G_1$ and $G_2$) while avoiding an
obstacle ($Obs$). The robot trajectories $\vx$ are now random due to the disturbances so we aim to achieve the robot's task with probability no less than $1-\delta$. We consider the chance constraint with $g(\vx,\vw)=\rho^\varphi(\vx)$ for
$
\varphi=F_{[2,6]} (G_1 \wedge F_{[3,7]} G_2) \wedge G_{[0,15]} (\neg Obs).
$
The cost function $J$ here is $\sum_{t=0}^{T-1} |\vu_t|$.

\paragraph{Task-Based Coordination using Capability Temporal Logic Planning}

We consider a CaTL planning problem over an environment $E = (Q,E,W,AP,L)$ with a randomly varying number of regions $N_r$ arranged in a grid, with adjacent regions connected with a constant edge weight of 2. The set of atomic propositions matches the set of regions and $L(q) = q$ for each $q \in Q$. A team of $N_a$ agents, each with one capability drawn from set $Cap$ with size $N_c$, must satisfy the following CaTL specification: 
$\varphi = \bigvee_{i=1}^{N_f} S_i,$
where
$S_i = (F_{[0,t]} (F_{[0,t]} T_0 \lor F_{[0,t]}((F_{[0,t]} T_1 \lor F_{[0,t]}(... \lor ((F_{[0,t]} T_{R-1} \lor F_{[0,t]}T_R))))) )$. Each task
$\mathcal{T}_i = \mathcal{T}(d_i, q_i, \{c_i, m_i\})$ has $d_i=1$ and $q_i,c_i,$ and $m_i$ drawn uniformly from $Q$, $Cap$, and $\{1,2,3, 4\}$, respectively. The cost function is the availability robustness of the plan, which is described in Section VI.B of \cite{leahy2021scalable}. In words, the availability robustness measures how robust the plan is to agent attrition.

\section{Experiments}
The two solver-guidance tasks require different difficulty regimes. Backdoor selection is evaluated on instances that Gurobi can usually solve to optimality within the cutoff, permitting runtime comparisons. Configuration is evaluated on harder instances for which SCIP often cannot close the gap; solution quality and progress over time are therefore the relevant outcomes. Table~\ref{tab:benchmarks} summarizes the corresponding benchmark sets. 

For each domain, we use 200 training instances and 100 held-out test instances. Experiments are conducted on 2.4 GHz Xeon-2640v3 CPUs with 64 GB memory. Training is performed on an NVIDIA V100 GPU with 112 GB memory. Models use a graph embedding width of 64, eight attention heads, Adam with learning rate $10^{-4}$, batch size 32, and up to 1,000 epochs; the checkpoint with the best validation loss is selected. We use Gurobi 11.0.0 for backdoor experiments and SCIP 9.0.0 for configuration experiments. 

\begin{table*}[!t]
\centering
\caption{\textbf{Solver configuration:} Methods are evaluated at a 900-second cutoff. PG and PI are computed against a per-instance reference value obtained from a 1,800-second Gurobi solve. ``Wins'' counts strict per-instance minima. Parentheses show reductions from Default SCIP. Lower values are better; bold marks the best mean and win count.}
\label{tab:config}
\begin{tabular}{llrrrrrr}
\toprule
\multirow{2}{*}{Benchmark} & \multirow{2}{*}{Method}
& \multicolumn{3}{c}{Primal Gap (\%)} & \multicolumn{3}{c}{Primal Integral}\\
\cmidrule(lr){3-5}\cmidrule(lr){6-8}
& & Wins & Mean & Std. Dev. & Wins & Mean & Std. Dev.\\
\midrule
\multirow{4}{*}{STL}
& Default SCIP & 8 & 25.15 & 37.30 & 13 & 293 & 242\\
& SMAC3-Transfer & 17 & 15.19 ($-40\%$) & 27.90 & 19 & 243 ($-17\%$) & 220\\
& Config-CL & 24 & 12.18 ($-52\%$) & \textbf{23.83} & 29 & 200 ($-32\%$) & \textbf{187}\\
& \textbf{DA-Config (ours)} & \textbf{37} & \textbf{10.95 ($-56\%$)} & 25.44 & \textbf{38} & \textbf{186 ($-37\%$)} & 206\\
\midrule
\multirow{4}{*}{CPP}
& Default SCIP & 5 & 43.30 & 64.28 & 0 & 21,111 & 10,746\\
& SMAC3-Transfer & 17 & 4.10 ($-91\%$) & 23.42 & 27 & 8,503 ($-60\%$) & \textbf{6,156}\\
& Config-CL & 20 & 2.74 ($-94\%$) & 12.99 & 31 & 8,381 ($-60\%$) & 6,979\\
& \textbf{DA-Config (ours)} & \textbf{32} & \textbf{2.25 ($-95\%$)} & \textbf{7.97} & \textbf{34} & \textbf{8,310 ($-61\%$)} & 6,175\\
\midrule
\multirow{4}{*}{CaTL}
& Default SCIP & 15 & 34.00 & 46.81 & 13 & 14,528 & 12,165\\
& SMAC3-Transfer & 17 & 33.33 ($-2\%$) & 46.47 & 19 & 14,167 ($-2\%$) & 11,792\\
& Config-CL & 22 & 32.57 ($-4\%$) & 45.54 & 27 & 13,560 ($-7\%$) & 12,346\\
& \textbf{DA-Config (ours)} & \textbf{40} & \textbf{28.99 ($-15\%$)} & \textbf{44.93} & \textbf{29} & \textbf{13,172 ($-9\%$)} & \textbf{11,775}\\
\bottomrule
\end{tabular}
\end{table*}

\paragraph{Backdoor baselines and metric.}
We compare Default Gurobi, uniform random branching priorities (Random), top-$k$ root-LP fractionality (LP-Frac), an MCTS-discovered backdoor transferred from training (MCTS-Transfer), generic Backdoor-Rank, and DA-Backdoor. Runtime is capped at 900 seconds; lower is better.

\paragraph{Configuration baselines and metrics.}
We compare Default SCIP, SMAC3-Transfer (a consensus configuration distilled by mode/median from the top five SMAC3 trials per training instance), Config-CL, and DA-Config under a 900-second limit. The primal gap is
$\mathrm{PG}=|v-v^*|/\max(|v^*|,10^{-8})$ when an incumbent $v$ exists and $vv^*\geq0$ \cite{berthold2006primal}; $v^*$ is obtained by a 1,800-second Gurobi run. The primal integral (PI) integrates PG over time \cite{achterberg2012rounding}. Lower values are better.

\subsection{Results}
\paragraph{Backdoor selection.}
Table~\ref{tab:backdoor} shows that DA-Backdoor has the lowest mean runtime in every domain. It improves over Default Gurobi by 20.1\% on STL, 14.4\% on CPP, and 20.4\% on CaTL. Relative to the Backdoor-Rank model, adding domain features reduces mean runtime by 10.4\%, 3.4\%, and 6.5\%, respectively. The reductions are not driven solely by a few easy instances: DA-Backdoor also obtains the lowest 75th percentile in all three domains and the lowest or tied-lowest median. Its variance is lower than the generic learned model in each domain, with the clearest reduction on CaTL (17.4 versus 23.2 seconds).

Performance nevertheless varies by statistic. Random records the most wins on STL, where many instances are solved quickly and small runtime differences can determine a win. DA-Backdoor has only 12 STL wins despite the lowest mean and median, indicating that its aggregate advantage comes from avoiding expensive runs rather than winning the largest number of short races. On CPP, Backdoor-Rank and DA-Backdoor tie at the median, and the domain-aware gain appears primarily in the upper half of the runtime distribution. The transferred and fractionality-based baselines degrade sharply on CPP, whereas both learned rankers improve on Default. On CaTL, DA-Backdoor leads both mean runtime and wins. Thus, the evidence supports a consistent improvement in mean and tail runtime, but not dominance on every instance or statistic.

\paragraph{Solver configuration.}
As shown in Table~\ref{tab:config}, DA-Config has the lowest mean PG and PI in each domain. Relative to Default SCIP, it reduces mean PG by 56\%, 95\%, and 15\% on STL, CPP, and CaTL, and mean PI by 37\%, 61\%, and 9\%. It also improves both reported means over Config-CL and SMAC3-Transfer in every domain. Because Config-CL and DA-Config differ only in their input features, this comparison provides the most direct evidence for the value of domain metadata.

The strongest gain occurs on CPP: the mean PG falls from 43.30 for Default SCIP to 2.25, and the PI falls from 21,111 to 8,310. Here the sample identity exposed by the metadata directly describes the replicated scenario structure that generic MILP features omit. STL also shows a substantial benefit, consistent with formula role, time, and depth distinguishing variables that otherwise have similar coefficient patterns. Gains on CaTL are smaller but remain consistent in both metrics. DA-Config records the most PG wins and PI wins in all domains. Standard deviations  uniformly improve on most domains, so the result should be not only interpreted as better average solution progress but also uniformly lower variability.

\section{Related Work}
\paragraph{MILP-based Planning.}
MILP encodings support planning and control under linear and signal temporal logic \cite{karaman2008optimal,raman2014model}, robust and multi-agent specifications \cite{sadraddini2015robust,liu2017communication}, and chance constraints \cite{calafiore2012mixed,zhao2024conformal}. Prior acceleration methods often predict binary assignments or active constraints \cite{masti2019learning,cauligi2021coco,bertsimas2022online}, whereas our approach leaves the formulation intact and learns to guide solver decisions. It is complementary to tighter or smaller encodings \cite{kurtz2022mixed}.

\paragraph{ML for General MILP Solving. }
A large body of work applies ML to guide algorithmic decisions in
Branch-and-Bound, including node selection
\citep{he2014learning,labassi2022learning}, variable branching
\citep{khalil2016learning,gasse2019exact,zarpellon2021parameterizing}, and
cut generation \citep{tang2020reinforcement,paulus2022learning,huang2022learning}.
Beyond BnB, ML has also been used to guide MILP meta-heuristics, such as
local search \citep{song2020general,sonnerat2021learning,huang2023searching,
tong2024optimization,cai2024balans}, and to predict high-quality partial
solutions using supervised learning
\citep{nair2020solving,han2022gnn,huang2024contrastive}. Recently, \citep{cai2025id} explore identity aware feature augmentation in the large-scale industry problems.
For complete surveys, see \citep{scavuzzo2024machine,huang2024distributional}.

\section{Conclusion}
We presented a domain-aware ML4CO framework for accelerating MILP-based motion planning with temporal logic and chance constraints. The framework exposes semantic structure from STL, CPP, and CaTL encodings to GNN policies for backdoor selection and solver configuration. On the tested distributions, DA-Backdoor attains the lowest mean runtime among six methods in all three domains, and DA-Config attains the lowest mean primal gap and primal integral among four methods in all three domains. Comparisons with matched generic-feature models indicate that domain metadata provides additional value beyond the MILP coefficient graph. Future work should isolate individual feature contributions, quantify uncertainty and total computational cost, test transfer across planning families, and extend domain-aware guidance to online and adaptive solver decisions. 

\section{Acknowledgments}
This work was supported in part by the National Science Foundation under grant numbers 2112533 (NSF Artificial Intelligence Research Institute for Advances in Optimization, AI4OPT), SHF-2048094 (CAREER Award), and IIS-SLES-2417075. Additional support was provided by Toyota R\&D and Siemens Corporate Research through the USC Center for Autonomy and AI, as well as by Ford Motors, Northrop Grumman, and an Amazon Faculty Research Award. We are thankful for Yiqi Zhao's help with the Multi-Target STL Planning case study.

DISTRIBUTION STATEMENT A. Approved for public release. Distribution is unlimited.
This material is based upon work supported by the Under Secretary of War for Research and Engineering under Air Force Contract No. FA8702-15-D-0001 or FA8702-25-D-B002. Any opinions, findings, conclusions or recommendations expressed in this material are those of the author(s) and do not necessarily reflect the views of the Under Secretary of War for Research and Engineering.
© 2026 Massachusetts Institute of Technology.
Delivered to the U.S. Government with Unlimited Rights, as defined in DFARS Part 252.227-7013 or 7014 (Feb 2014). Notwithstanding any copyright notice, U.S. Government rights in this work are defined by DFARS 252.227-7013 or DFARS 252.227-7014 as detailed above. Use of this work other than as specifically authorized by the U.S. Government may violate any copyrights that exist in this work.

%

\appendix



\bibliography{sample}
\newpage


\onecolumn
\section{Detailed Planning Formulations}
\label{app:planning}

\subsection{DT-STL Motion Planning and MILP Encoding}
\label{app:STL}

The discrete-time Signal Temporal Logic (DT-STL) planning problem is
\begin{subequations}
\label{eq:STL_oc}
\begin{align}
    \min_{\vx,\vu}\quad & J(\vx,\vu)\\
    \text{s.t.}\quad
    & \vx_{t+1}=f_t(\vx_t,\vu_t),
      && t=0,\ldots,T-1,\\
    & \vx_0\in X_0,\\
    & (\vx,0)\models\varphi.
    \label{eq:STL_constraint}
\end{align}
\end{subequations}
This problem differs from the generic MILP-based planning problem in
\eqref{eq:milp_oc} through the final satisfaction constraint.

We use \texttt{stlpy} \citep{kurtz2022mixed}, which implements the MILP
encoding described in \citep[Section~5]{belta2019formal}. The encoding
assumes that $\varphi$ is in negative normal form (NNF). This is without
loss of generality because every DT-STL formula can be transformed into a
semantically equivalent NNF formula $\varphi_{\mathrm{NNF}}$
\citep[Section~4]{sadraddini2015robust}.

For each predicate $h(\vx_t)\geq 0$, a binary variable
$z_t^h\in\{0,1\}$ represents predicate satisfaction. A standard Big-M
encoding at threshold $\rho_{\min}\geq 0$ is
\begin{align}
    h(\vx_t) &\geq \rho_{\min}-M(1-z_t^h),\\
    h(\vx_t) &\leq \rho_{\min}-\epsilon+Mz_t^h,
\end{align}
where $M>0$ is sufficiently large and $\epsilon>0$ is a numerical
tolerance. Thus, $z_t^h=1$ indicates that the predicate is satisfied with
margin at least $\rho_{\min}$.

Boolean operators are encoded recursively. For example, if
$z_t^\psi$ represents $\psi=\varphi_1\wedge\varphi_2$, then
\begin{align}
    z_t^\psi &\leq z_t^{\varphi_1},&
    z_t^\psi &\leq z_t^{\varphi_2},\\
    z_t^\psi &\geq
    z_t^{\varphi_1}+z_t^{\varphi_2}-1.
\end{align}
If $\psi=\varphi_1\vee\varphi_2$, then
\begin{align}
    z_t^\psi &\geq z_t^{\varphi_1},&
    z_t^\psi &\geq z_t^{\varphi_2},\\
    z_t^\psi &\leq
    z_t^{\varphi_1}+z_t^{\varphi_2}.
\end{align}
Eventually, always, and until operators are encoded by applying analogous
Boolean constraints over their corresponding time intervals; see
\citep{belta2019formal} for the complete construction. Formula
satisfaction is enforced by
\begin{equation}
    z_0^{\varphi_{\mathrm{NNF}}}=1.
\end{equation}
This root constraint corresponds to $c(\vx,\vz)=1$ in
\eqref{eq:milp_oc}. The semantic roles, time indices, and formula depths
created during this recursive construction form part of the metadata used
by DA-Backdoor and DA-Config.

\subsection{Chance-Constrained Planning via CPP}
\label{app:Chance}

Consider dynamics affected by a random disturbance:
\begin{equation}
    \vx_{t+1}=f_t(\vx_t,\vu_t,\vw_t),
\end{equation}
where $\vw=(\vw_0,\ldots,\vw_{T-1})$ has an unknown or analytically
intractable distribution. The chance-constrained planning problem is
\begin{subequations}
\label{eq:chance_oc}
\begin{align}
    \min_{\vx,\vu}\quad & J(\vx,\vu)\\
    \text{s.t.}\quad
    & \vx_{t+1}=f_t(\vx_t,\vu_t,\vw_t),
      &&t=0,\ldots,T-1,\\
    & \vx_0\in X_0,\\
    & \operatorname{Prob}\!\left(g(\vx,\vw)\geq 0\right)
      \geq 1-\delta,
    \label{eq:chance_constraint}
\end{align}
\end{subequations}
where $\delta\in(0,1)$ is the allowed failure probability. In our timed
reach-avoid benchmark, $g$ is the STL robustness
$\rho^\varphi(\vx)$.

Conformal Predictive Programming (CPP) approximates
\eqref{eq:chance_constraint} using $K$ sampled disturbance trajectories
$\{\vw^{(i)}\}_{i=1}^K$ \citep{zhao2024conformal}. The control sequence is
shared across samples, while each sample induces a separate state
trajectory $\vx^{(i)}$:
\begin{subequations}
\label{eq:CPP_oc}
\begin{align}
    \min_{\vu,\{\vx^{(i)}\}_{i=1}^K}\quad & J(\vu)\\
    \text{s.t.}\quad
    & \vx_{t+1}^{(i)}
      =f_t(\vx_t^{(i)},\vu_t,\vw_t^{(i)}),
      &&\substack{t=0,\ldots,T-1,\\i=1,\ldots,K},\\
    & \vx_0^{(i)}=\vx_0\in X_0,
      &&i=1,\ldots,K,\\
    & \sum_{i=1}^K
      \mathds{1}\!\left\{
      g(\vx^{(i)},\vw^{(i)})\geq 0
      \right\}
      \geq
      \left\lceil (K+1)(1-\delta)\right\rceil.
    \label{eq:CPP_quantile}
\end{align}
\end{subequations}

To obtain an MILP, we introduce a binary variable $z_i$ for each sample.
For sufficiently large $M$, sample satisfaction can be enforced through
\begin{equation}
    g(\vx^{(i)},\vw^{(i)})\geq -M(1-z_i),
    \qquad i=1,\ldots,K,
\end{equation}
together with
\begin{equation}
    \sum_{i=1}^K z_i
    \geq \left\lceil (K+1)(1-\delta)\right\rceil.
\end{equation}
Additional inequalities may be included when exact equivalence between
$z_i=1$ and sample satisfaction is required. The sample identities,
time indices, and distinction between shared and sample-specific
variables are retained as domain-aware metadata. See
\citep[Chapter~5]{zhao2024conformal} for the complete CPP construction.

\subsection{Capability Temporal Logic Planning}
\label{app:CaTL}

Capability Temporal Logic (CaTL) is designed for coordinating
heterogeneous agents in a partitioned environment
\citep{leahy2021scalable}. Let
\begin{equation}
    Env=(Q,E,W,AP,L),
\end{equation}
where $Q$ is a finite set of regions, $E\subseteq Q\times Q$ is the set
of directed transitions, $W:E\rightarrow\mathbb{N}$ gives transition
times, $AP$ is a set of region propositions, and
$L:Q\rightarrow 2^{AP}$ labels the regions.

Let $Cap$ be the set of capabilities and $\mathcal{J}$ the agent index
set. Agent $j\in\mathcal{J}$ is represented by
\begin{equation}
    A_j=(q_{0,j},Cap_j),
\end{equation}
where $q_{0,j}\in Q$ is its initial region and $Cap_j\in Cap$ is its
capability. Self-loops with $W(q,q)=1$ allow agents to remain in a region.

For each capability $c$, region $q$, edge $e$, and time $t$, define
\begin{align}
    x_t^{q,c}
    &:= \text{number of agents with capability $c$ in region $q$},\\
    u_t^{e,c}
    &:= \text{number of agents with capability $c$ entering edge $e$}.
\end{align}
The initial occupancies are
\begin{equation}
\label{eq:CaTLInitCond}
    x_0^{q,c}
    =
    \sum_{j\in\mathcal{J}}
    \mathds{1}\{q_{0,j}=q\}
    \mathds{1}\{Cap_j=c\}.
\end{equation}
For $\operatorname{In}(q)=\{e=(q',q)\in E\}$ and
$\operatorname{Out}(q)=\{e=(q,q')\in E\}$, delayed flow conservation is
encoded by
\begin{align}
    x_t^{q,c}
    &=
    \sum_{e\in\operatorname{In}(q)}
    u_{t-W(e)}^{e,c},
    \label{eq:CaTLOccupancy}\\
    \sum_{e\in\operatorname{Out}(q)}u_t^{e,c}
    &=x_t^{q,c},
    \label{eq:CaTLFlow}
\end{align}
where $u_\tau^{e,c}=0$ for $\tau<0$. These equations provide the
integer-valued team dynamics used in the CaTL MILP.

A counting proposition $cp_i=(c_i,m_i)$ requires at least $m_i$ agents
with capability $c_i$. A CaTL task is
\begin{equation}
    \mathcal{T}
    =(d,\pi,\{cp_i\}_{i\in\mathcal{I}_{\mathcal{T}}}),
\end{equation}
where $d$ is the task duration and $\pi\in AP$ identifies the relevant
regions. The task is satisfied when all capability requirements hold in
the regions labeled by $\pi$ for $d$ consecutive time steps.

The syntax considered in this paper is
\begin{equation}
\label{eq:catlfrag_app}
    \varphi
    ::= \top
    \mid \mathcal{T}
    \mid \neg\varphi
    \mid \varphi_1\wedge\varphi_2
    \mid \varphi_1\until_{\intvl}\varphi_2.
\end{equation}
As with STL, derived operators include disjunction, eventually, and
always. CaTL also admits an availability-robustness semantics that
measures the number of agents that can be removed while preserving
formula satisfaction \citep{leahy2021scalable}.

The CaTL planning problem is
\begin{subequations}
\label{eq:CaTL_oc}
\begin{align}
    \min_{\vx,\vu,\vz}\quad & J(\vx,\vu)\\
    \text{s.t.}\quad
    & \text{initial conditions \eqref{eq:CaTLInitCond}},\\
    & \text{team dynamics
       \eqref{eq:CaTLOccupancy}--\eqref{eq:CaTLFlow}},\\
    & (\vx,0)\models\varphi.
\end{align}
\end{subequations}

For illustration, let $z_t^{q,cp_i}\in\{0,1\}$ indicate whether region
$q$ contains at least $m_i$ agents with capability $c_i$ at time $t$.
This relation is encoded using Big-M inequalities. Let
$z_t^{\pi,cp_i}$ indicate that the requirement holds in every region
labeled by $\pi$. It is encoded as
\begin{align}
    z_t^{\pi,cp_i}
    &\leq z_t^{q,cp_i},
    &&q\in L^{-1}(\pi),\\
    z_t^{\pi,cp_i}
    &\geq
    \sum_{q\in L^{-1}(\pi)}z_t^{q,cp_i}
    -|L^{-1}(\pi)|+1.
\end{align}
Let $z_t^{\pi,\mathcal{I}_{\mathcal{T}}}$ indicate that all capability
requirements hold:
\begin{align}
    z_t^{\pi,\mathcal{I}_{\mathcal{T}}}
    &\leq z_t^{\pi,cp_i},
    &&i\in\mathcal{I}_{\mathcal{T}},\\
    z_t^{\pi,\mathcal{I}_{\mathcal{T}}}
    &\geq
    \sum_{i\in\mathcal{I}_{\mathcal{T}}}z_t^{\pi,cp_i}
    -|\mathcal{I}_{\mathcal{T}}|+1.
\end{align}
Finally, task satisfaction over duration $d$ is represented by
$z_t^{\mathcal{T}}$:
\begin{align}
    z_t^{\mathcal{T}}
    &\leq z_\ell^{\pi,\mathcal{I}_{\mathcal{T}}},
    &&\ell=t,\ldots,t+d-1,\\
    z_t^{\mathcal{T}}
    &\geq
    \sum_{\ell=t}^{t+d-1}
    z_\ell^{\pi,\mathcal{I}_{\mathcal{T}}}-d+1.
\end{align}
The remaining Boolean and temporal operators follow the recursive STL
encoding. Capability identities, region identities, time indices, and
semantic indicator roles are retained in the metadata used by the
domain-aware models.

\section{Solver Configuration Parameters}
\label{app:config}

The Config-CL and DA-Config experiments tune the following 15 SCIP
parameters. They cover branching, cut management, LP processing, and
node selection.

\begin{itemize}
    \item \textbf{branching/clamp}: Minimum relative distance between a
    continuous branching point and the variable bounds.

    \item \textbf{branching/lpgainnormalize}: Normalization strategy for
    LP gains when updating continuous-variable pseudocosts.

    \item \textbf{branching/midpull}: Fraction by which a continuous
    branching point is shifted toward the midpoint of its domain.

    \item \textbf{branching/midpullreldomtrig}: Relative-domain threshold
    used to scale \textit{midpull}.

    \item \textbf{branching/preferbinary}: Whether binary variables are
    preferred as branching candidates.

    \item \textbf{branching/scorefac}: Weight assigned to downward and
    upward gain predictions in the branching score.

    \item \textbf{branching/scorefunc}: Branching score aggregation rule
    (\textit{sum}, \textit{product}, or \textit{quotient}).

    \item \textbf{cutselection/hybrid/minortho}: Minimum orthogonality
    required for a cut to enter the LP.

    \item \textbf{cutselection/hybrid/minorthoroot}: Minimum
    orthogonality required for a cut to enter the root LP.

    \item \textbf{lp/colagelimit}: Maximum age of a dynamic column before
    it is removed from the LP.

    \item \textbf{lp/pricing}: LP pricing strategy.

    \item \textbf{lp/rowagelimit}: Maximum age of a dynamic row before it
    is removed from the LP.

    \item \textbf{nodeselection/childsel}: Rule for selecting the child
    node explored next.

    \item \textbf{separating/cutagelimit}: Maximum age of a cut before
    removal from the global cut pool.

    \item \textbf{separating/poolfreq}: Frequency at which the global cut
    pool is separated.
\end{itemize}

\section{Benchmark Generation Details}
\label{app:instances}

The backdoor-selection and solver-configuration experiments use different
difficulty regimes, as summarized in Table~\ref{tab:benchmarks}. This
section gives additional details about the instance generators.

\subsection{Multi-Target STL Planning}

We use \texttt{stlpy} \citep{kurtz2022mixed} to construct the STL MILPs.
Targets and obstacles are placed randomly in a two-dimensional workspace.
The robot follows double-integrator dynamics and must reach at least one
target from each target group while avoiding every obstacle over the
planning horizon.

The backdoor-selection benchmark uses
\begin{equation}
    (N_o,N_c,N_t,T)=(2,5,2,30),
\end{equation}
where $N_o$ is the number of obstacles, $N_c$ is the number of target
groups, $N_t$ is the number of targets per group, and $T$ is the planning
horizon. The harder solver-configuration benchmark uses
\begin{equation}
    (N_o,N_c,N_t,T)=(3,5,3,35).
\end{equation}

\subsection{Chance-Constrained Timed Reach-Avoid Planning}

The CPP benchmark uses a two-dimensional robot with noisy
double-integrator dynamics. At each time step, position and velocity are
perturbed by zero-mean Gaussian noise with covariance
$0.01\mathbb{I}_4$. Bounds are imposed on acceleration and velocity.

We sample $K=20$ disturbance trajectories and use a planning horizon of
$T=15$. The same parameter setting is used for both solver-guidance
tasks. A shared control sequence is optimized across samples, and at
least
\begin{equation}
    \left\lceil(K+1)(1-\delta)\right\rceil
\end{equation}
sampled trajectories must satisfy the timed reach-avoid specification.

\subsection{Capability Temporal Logic Coordination}

The CaTL environment contains five capability labels and five regions.
Agents are initialized randomly, and their capabilities are sampled from
the available capability set. Adjacent regions are connected with
transition time two.

Specifications are generated recursively from CaTL tasks, disjunctions,
and eventually operators. Each task has duration one, while its region,
required capability, and required number of agents are sampled randomly.
The backdoor-selection benchmark uses horizon $T=42$, whereas the harder
solver-configuration benchmark uses $T=50$. Thus, the reported benchmark
parameters are $(5,5,42)$ and $(5,5,50)$, respectively, as shown in
Table~\ref{tab:benchmarks}.

\section{Additional Experimental Details}
\label{app:experiments}

\subsection{DA-Backdoor Training Data}

For Backdoor-Rank and DA-Backdoor, we use the Monte Carlo Tree Search
(MCTS) procedure of \citet{khalil2022finding} to generate candidate
backdoor sets for each training instance. Every candidate is evaluated
using the default Gurobi configuration, and its time to reach optimality
is recorded.

We retain the 15 fastest and 15 slowest candidates and form ranking pairs
between these groups. Backdoor-Rank and DA-Backdoor use the same
candidate sets, labels, architecture, and optimization settings; they
differ only in whether domain-aware metadata are included.

At test time, we sample 50 candidate sets of size eight from integer
variables that are fractional in the root LP relaxation. The GAT scores
each candidate, and the highest-ranked set is assigned elevated branching
priority during BnB. The learned model changes only branching priorities;
the underlying Gurobi search and optimality guarantees remain unchanged.

\subsection{DA-Config Training Data}

For Config-CL and DA-Config, SMAC3 \citep{lindauer2022smac3} generates
candidate settings over the 15 SCIP parameters listed in
Appendix~\ref{app:config}. Configurations producing the best feasible
solutions within the labeling budget form the positive set, while those
producing the worst feasible solutions form the negative set.

The GAT is trained using the contrastive objective in
\eqref{eq:contrastive}. At inference, the graph-level representation is
decoded into a valid SCIP configuration and applied before solving.
Config-CL and DA-Config again use identical training instances,
configuration labels, architectures, and optimization settings; only the
domain-aware metadata differ.

\subsection{Implementation Details}

For each planning domain, we use 200 training instances and 100 held-out
test instances. All solver runs are performed on 2.4\,GHz Xeon-2640v3
CPUs with 64\,GB of memory. Model training is performed using an NVIDIA
V100 GPU.

The GAT uses an embedding dimension of $L=64$ and eight attention heads.
Models are trained with Adam \citep{kingma2014adam}, learning rate
$10^{-4}$, batch size 32, and at most 1,000 epochs. The checkpoint with
the lowest validation loss is used for evaluation.

We use Gurobi~11.0.0 for the backdoor-selection experiments and
SCIP~9.0.0 for the solver-configuration experiments. Backdoor sets have
size eight, following \citet{cai2024learning}. All reported test runs use
a 900-second cutoff. The solver-configuration reference objective is
obtained from a separate 1,800-second Gurobi run, as described in the
main text.

\end{document}

%% file: macros.tex
\newcommand{\alw}{\mathbf{G}}
\newcommand{\ev}{\mathbf{F}}
\newcommand{\until}{\mathbf{U}}

\newcommand{\intvl}{\mathcal{I}}

\newcommand{\vx}{\mathbf{x}}
\newcommand{\vu}{\mathbf{u}}
\newcommand{\vw}{\mathbf{w}}
\newcommand{\vz}{\mathbf{z}}
\newcommand{\horizon}{\mathsf{hrz}}